\documentclass[aps,showpacs,preprint,preprintnumber,nofootinbib,amsmath,amssymb,ascmac,bm,12pt]{revtex4}
\usepackage{bm}
\usepackage[dvipdfmx]{graphicx,color}

\begin{document}
\title{\vspace*{0.5cm}
Charged Black Holes in a Five-dimensional Kaluza-Klein Universe}
\bigskip
\author{
${}^{1}$Yuki Kanou,
${}^{1}$Hideki Ishihara\footnote{E-mail: ishihara@sci.osaka-cu.ac.jp}, 
${}^{2}$Masashi Kimura\footnote{E-mail: m.kimura@damtp.cam.ac.uk},  
${}^{1}$Ken Matsuno\footnote{E-mail: matsuno@sci.osaka-cu.ac.jp},  
and
${}^{1}$Takamitsu Tatsuoka\footnote{E-mail: tatsuoka@sci.osaka-cu.ac.jp}
\bigskip
\bigskip
}
\affiliation{
${}^{1}$Department of Mathematics and Physics, Osaka City University, Sumiyoshi, Osaka 558-8585, Japan
\\
${}^{2}$DAMTP, University of Cambridge, Centre for Mathematical Sciences,
Wilberforce Road, Cambridge CB3 0WA, UK
\bigskip
\bigskip
}

\begin{abstract}
We examine an exact solution which represents a charged black hole in 
a Kaluza-Klein universe in the five-dimensional Einstein-Maxwell theory. 
The spacetime approaches to the five-dimensional Kasner solution that  
describes expanding three dimensions and shrinking an extra dimension 
in the far region. 
The metric is continuous but not smooth 
at the black hole horizon. 
There appears a mild curvature singularity that a free-fall observer can 
traverse the horizon.
The horizon is a squashed three-sphere with a constant size, 
and the metric is approximately static near the horizon. 
\end{abstract}

\preprint{OCU-PHYS-408}
\preprint{AP-GR-113}


\date{\today}
\maketitle

\section{Introduction}\label{intro}

Higher-dimensional spacetimes are investigated extensively in the context of 
unified theories. 
Spacetimes of the Kaluza-Klein type, non-compact three-dimensional space with  
compact extra dimensions of small size, 
would accord with effective four-dimensional spacetime. 
Black holes with compact extra dimensions, so-called Kaluza-Klein black holes, 
could be suitable for the model describing black holes that reside 
in our four-dimensional world. 
For example, five-dimensional squashed Kaluza-Klein black hole solutions 
\cite{Dobiasch:1981vh, Gibbons:1985ac, Gauntlett:2002nw, 
Gaiotto:2005gf, Ishihara:2005dp, Wang:2006nw, 
Yazadjiev:2006iv, Nakagawa:2008rm, Tomizawa:2008hw, Tomizawa:2008rh, Stelea:2008tt,  
Tomizawa:2008qr, Gal'tsov:2008sh, Matsuno:2008fn, Bena:2009ev, Tomizawa:2010xq, 
Mizoguchi:2011zj, Chen:2010ih, Stelea:2011fj, Nedkova:2011hx, Nedkova:2011aa, 
Tatsuoka:2011tx, Mizoguchi:2012vg, 
Matsuno:2012hf, Matsuno:2012ge, Tomizawa:2012nk, Stelea:2012ph, Nedkova:2012yn, 
Wu:2013nea, Brihaye:2013vsa}
behave as fully five-dimensional black holes near the horizon 
and asymptote to the four-dimensional Minkowski spacetime with a compact extra dimension. 
The squashed black holes are constructed on the Gross-Perry-Sorkin (GPS) 
monopole solution \cite{Gross:1983, Sorkin:1983ns}, 
and they have smooth horizons. 

One of the important questions for the Kaluza-Klein spacetime model is 
why the size of extra dimensions are too small to detect by experiments. 
An interesting explanation 
is that the three-dimensional space expands while the extra dimensions shrink enough 
in the history of the universe 
\cite{Chodos:1979vk, Sahdev:1988fp, Ishihara:1984wx, Ishihara:1986if}. 
In the five-dimensional case, the (4+1)-dimensional Kasner solution, 
three-dimensional space expands while an extra dimension shrinks with the time evolution, 
provides such a model universe. 
Gibbons, Lu and Pope generalized the GPS monopole solution 
to dynamical ones \cite{Gibbons:2005rt}, 
which nicely behaves as the Kasner universe in a distant region.

As generalizations of the Kastor-Traschen solution that describes charged black holes 
in de Sitter universe \cite{Kastor:1992nn}, 
there exist a lot of solutions on higher-dimensional black holes for the Einstein-Maxwell 
system \cite{
Maki:1992tq, Klemm:2000vn,Ishihara:2006ig,Ida:2007vi,Matsuno:2007ts, 
Gibbons:2007zu}, 
and brane solutions \cite{Gibbons:2005rt, Maeda:2009zi} 
in expanding universe models. 
All of these solutions are constructed using harmonic functions on four-dimensional Ricci flat 
spaces, where the harmonic functions contain the time-coordinate as a parameter. 
Time evolution of these spacetimes are driven by a cosmological constant, and then, 
all spatial dimensions expand in the laps of time. 

In this paper, 
we investigate black hole solutions on the dynamical GPS monopole solution derived 
by Gibbons, Lu and Pope \cite{Gibbons:2005rt}. Since a time slice of the dynamical 
GPS monopole solution is a Ricci flat space which has the time-variable as a parameter, 
by using suitable harmonic functions on the space 
we can construct an exact time-dependent solution in the five-dimensional 
Einstein-Maxwell theory. \footnote{
By the Kaluza-Klein reduction of the solution in this paper, 
we obtain a solution in Einstein-Maxwell-dilaton system discussed 
in ref. \cite{Maeda:2009ds, Maeda:2010ja}.
} 
The solution approaches to the dynamical GPS monopole solution 
in the far region, expanding three dimensions and shrinking a compact dimension, then 
it describes a charged black hole in a Kaluza-Klein universe. 
What happens to the black hole when the extra dimension shrinks so that its size  
becomes smaller than the size of the black hole? 
To answer this question, we study geometrical properties of the solution.

This paper is organized as follows.  
We present the explicit form of the solution in Sec.\ref{bhsols}. 
The curvature singularities are studied in Sec.\ref{singularities}. 
We give a C$^0$ extension of the metric, and show that the solution describes 
a black hole in Sec.\ref{eventhorizon}. 
Geometrical properties of the event horizon are also discussed. 
We summarize our results in Sec.\ref{summary}.

\section{Exact solution}\label{bhsols}

We consider exact charged dynamical solutions in the five-dimensional 
Einstein-Maxwell theory with the action\footnote{
The solution discussed in this paper is also a solution of 
the Einstein-Maxwell-Chern-Simons system.
}  
\begin{eqnarray}
 S = \frac{1}{16\pi} \int d^5 x \sqrt{-g} 
  \left( R - F_{\mu\nu} F^{\mu\nu} \right) .
\end{eqnarray}
The metric and the Maxwell field of the solutions are written as  
\begin{eqnarray}
 ds^2 &=&  -H^{-2} dt^2 + H \left[ V ( dr^2 + r^2 d\Omega^2_{\rm S^2} ) 
  + N^2 V^{-1} ( d\psi + \cos \theta d\phi )^2 \right], 
\label{eq:metric}
\\
 A_\mu dx^\mu &=&  \pm \frac{ \sqrt{3} }{ 2 } H^{-1} dt, 
\label{eq:field}
\end{eqnarray}
where $d\Omega^2_{\rm S^2} = d\theta^2 + \sin ^2 \theta d\phi^2$ is the metric of 
unit two-dimensional sphere, S$^2$,  
and the functions $H$ and $V$ are given by 
\begin{eqnarray}
 H &=&  1 + \frac{ M }{ r },
\label{eq:h}
\\
 V &=&  \frac{ t }{ N } + \frac{ N }{ r } ,
\label{eq:v}
\end{eqnarray}
where $M$ and $N$ are positive constants. 
As will be shown later, the metric describes black holes in an expanding Kaluza-Klein universe. 
The coordinates run the ranges of 
$-\infty < t < \infty , ~-M < r < \infty,~0 \leq \theta \leq\pi,~ 0 \leq \phi \leq 2\pi $, and $0 \leq \psi \leq 4\pi $. 
The angular part of the space consists 
twisted S$^1$ bundle over S$^2$.  

If the parameter $M$ vanishes,  
the solution \eqref{eq:metric} coincides with the vacuum dynamical GPS monopole 
solution \cite{Gibbons:2005rt} 
\begin{eqnarray}
 ds^2 &=&  - dt^2 
 + \left(\frac{ t }{ N } + \frac{ N }{ r }\right) ( dr^2 + r^2 d\Omega^2_{\rm S^2} ) 
  + N^2 \left(\frac{ t }{ N } + \frac{ N }{ r }\right)^{-1} 
  ( d\psi + \cos \theta d\phi )^2 . 
\label{eq:dyn_GPS}
\end{eqnarray}
It is easily seen by a coordinate transformation that the point $r=0$ on a 
time slice $t={\rm const.}$ is regular, and
there is an initial singularity at $tr=-N^2$. 
In the limit $r \to + \infty$ with $t ={\rm finite}$, 
the metric \eqref{eq:metric} as same as the metric \eqref{eq:dyn_GPS} approaches to  
the five-dimensional Kasner-like metric with twisted S$^1$ in the form 
\begin{eqnarray}
 ds^2 \simeq - dt^2 + \frac{ t }{ N } ( dr^2 + r^2 d\Omega^2_{\rm S^2} ) 
  + N^2 \frac{ N }{ t } ( d\psi + \cos \theta d\phi )^2, 
\label{eq:5dkasner}
\end{eqnarray}
where  
the size of three-dimensional space increases and the size of S$^1$ decreases 
as the time $t$ laps \cite{Bizon:2006ue}. 
It is clear that the metric \eqref{eq:metric} has a null infinity 
at $r = + \infty ,~ t = + \infty$ 
in the Kasner-like region. 
On the other hand, 
in the limit $r \to 0$ with $t ={\rm finite}$,  
the metric \eqref{eq:metric} approaches
\begin{eqnarray}
 ds^2 \simeq - \frac{ r^2 }{ M^2 } dt^2 + \frac{ MN }{ r^2 } dr^2 
 + MN \left[ d\Omega^2_{\rm S^2} + ( d\psi + \cos \theta d\phi )^2 \right] . 
\label{eq:AdS}
\end{eqnarray}
It will be clarified, the metric \eqref{eq:AdS} in the form of AdS$_2 \times$S$^{3}$ 
does not describe near horizon geometry. 

The black hole solution \eqref{eq:metric} is constructed on the dynamical GPS solution. 
If we take a Kaluza-Klein reduction of the metric \eqref{eq:metric} with respect to 
the Killing vector field $\partial / \partial \psi $, 
the obtained four-dimensional Einstein-Maxwell-dilaton solution coincides with that in 
\cite{Maeda:2009ds, Maeda:2010ja}.   

\section{Curvature singularities}\label{singularities}

\begin{figure}[h]
\begin{center}
 \includegraphics[width=110mm]{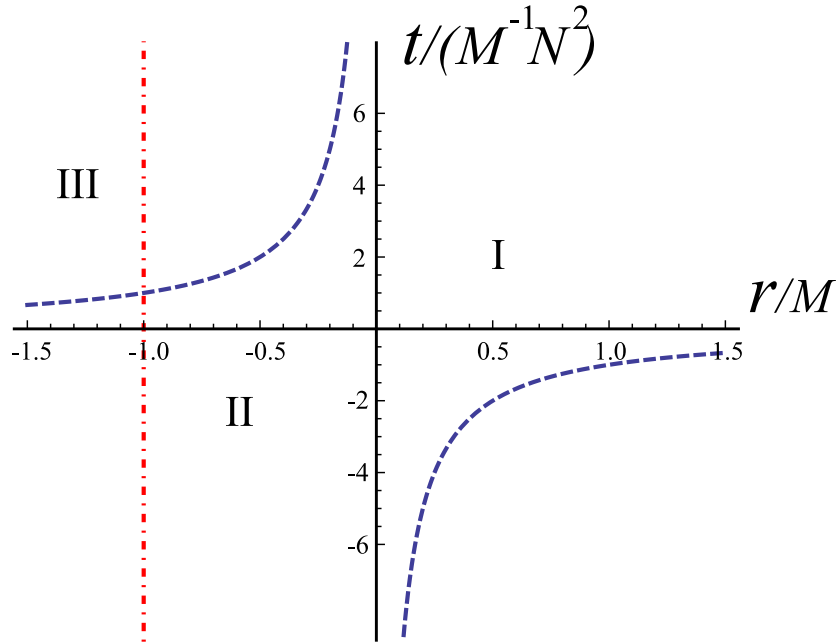}
\end{center}
\caption{Curvature singularities are shown by dashed (blue) curves, 
$t r = - N^2$, and a dot-dashed (red) line, $r = -M$, in the $t$-$r$ plane.
The metric \eqref{eq:metric} describes a spacetime with the signature $(-,+,+,+,+)$ 
in three regions: 
I ($r>0$, $rt>-N^2$), II ($-M<r<0$, $rt>-N^2$), and III ($r<-M$, $rt<-N^2$). 
}
\label{tokuitenzu}
\end{figure}

To examine the global structure of the spacetime with the metric \eqref{eq:metric}, 
we first seek the locations of curvature singularities.  
The Kretschmann invariant and the 
square of the Maxwell field are 
\begin{eqnarray}
  R^{\mu \nu \rho \sigma} R_{\mu \nu \rho \sigma} 
  &\propto &  \frac{ 1 }{ ( M + r )^6 ( N^2 + rt )^6 } ,
\\
  F^{\mu \nu } F_{\mu \nu } &\propto & \frac{ 1 }{ ( M + r )^3 ( N^2 + rt ) } .
\end{eqnarray}
We see that there are curvature singularities at $r = -M$ and 
$r t  = -N^2$.  
These two singularities intersect at $t =  N^2 / M > 0 ,~ r = -M < 0$ 
(see Fig.\ref{tokuitenzu}).

For the spacetime signature $(-,+,+,+,+)$, the inequality $H(r)V(t,r)>0$ should be hold. 
Namely, we can consider three regions: (I) $r>0$, $rt>-N^2$, (II) $-M<r<0$, $rt>-N^2$, 
and (III) $r<-M$, $rt<-N^2$. 
Since the metric \eqref{eq:metric} in the region III describes a spacetime 
with a naked singularity, then we concentrate on the regions I and II. We will show, 
hereafter, that 
the metric \eqref{eq:metric} in the combined regions I and II represents a black hole.

\bigskip

We consider the normal vector field 
$n^{(1)}_{\mu } dx^\mu = dr$ to the $r={\rm const.}$ surfaces. 
The norm of the vector field is given by
\begin{eqnarray}
 g^{\mu\nu}n^{(1)}_\mu n^{(1)}_{\nu} =  H^{-1} V^{-1}. 
\label{eq:norm1}
\end{eqnarray}
Since the norm is positive in the region II, 
then the curvature singularity $r=-M$ in the region II is timelike. 

We also consider the normal vector field 
$n^{(2)}_{\mu } dx^\mu = rdt +  tdr$ to the $rt={\rm const.}$ surfaces. 
The norm of the vector is   
\begin{eqnarray}
 g^{\mu\nu}n^{(2)}_\mu n^{(2)}_{\nu} = -H^2 r^2 + H^{-1} V^{-1} t^2  
 \label{eq:norm}
\end{eqnarray}
In the limit $rt \to -N^2$ in the regions I and II, 
the norm \eqref{eq:norm} becomes $+\infty$. 
Then, the curvature singularities $rt = -N^2$ in the both region I and II are timelike.

\section{Extension of Spacetime} \label{eventhorizon}

\subsection{New Coordinates} \label{utnullcoord}

The metric \eqref{eq:metric} has an apparent singularity at $r=0$. 
We investigate the possibility of extension by using null geodesics starting 
from the region $r>0$. 

{
If we restrict ourselves to the null geodesics confined in the $t$-$r$ plane, i.e., 
$\theta ={\rm const.}$, $\phi ={\rm const.}$, $\psi ={\rm const.}$, 
the null geodesics are determined by the null condition, 
\begin{eqnarray}
 - H^{-2} dt ^2 + H V dr^2 = 0 . 
\label{eq:nullcond} 
\end{eqnarray}
}
Then, we have 
\begin{eqnarray}
 \left( \frac{dt}{dr} \right) ^2 = \frac{(r+M)^3 (tr + N^2)}{N r^4} .
\label{eq:nullcond2} 
\end{eqnarray} 
For ingoing future null geodesics, increasing $t$ and decreasing $r$, we see that 
$t$ should diverge as $r\to 0$ with $tr=\text{finite}$. The null geodesics terminate 
the coordinate boundary $r\to 0$ with $tr=\text{finite}$. 
Then, we try to extend the metric there. 

In many cases, a set of null geodesics is a powerful tool to construct coordinates covering 
the black hole horizons. Unfortunately, in our case, we hardly solve \eqref{eq:nullcond2} in 
analytic form. 
Then, we use curves that are approximate solutions for \eqref{eq:nullcond2} in the vicinity of $r = 0$.   
We assume the approximate solutions in the form, 
\begin{eqnarray}
  tr = a + u r^b + c r ,
\label{eq:nullcondsolassume} 
\end{eqnarray} 
where $a ,~ b,~ c$ are constants, and $u$ is an arbitrary parameter. 
Substituting \eqref{eq:nullcondsolassume} into \eqref{eq:nullcond2}, 
and taking the limit $r\to 0$, we can determine the constants $a ,~ b,~ c$ 
as
\begin{eqnarray}
 a &=&  \frac{M^3}{2N}\left(1 + \sqrt{1+\frac{4 N^3}{M^3}}\right) ,
\label{eq:nullcondsolassumea}
 \\
 b &=&  1 - \frac{M^3}{2 a N} 
 =1-\frac{1}{1 + \sqrt{1+4 N^3/M^3}}, 
\label{eq:nullcondsolassumeb}
 \\
 c &=&  - \frac{3 (a + N^2)}{M} 
 = -\frac{3M^2}{2N} \left( 1+\frac{2N^3}{M^3} +\sqrt{1+\frac{4 N^3}{M^3}}\right). 
\label{eq:nullcondsolassumec}
\end{eqnarray} 
The constant $b$ takes a value in the range $1/2 < b < 1$. 

The curves \eqref{eq:nullcondsolassume} with \eqref{eq:nullcondsolassumea}-\eqref{eq:nullcondsolassumec}
are approximately ingoing future null geodesics that attain the coordinate boundary.  
The free parameter $u$, which labels the curves, can be used as a new coordinate. 

Now, we introduce new coordinates $(\rho ,~ u)$ as 
\begin{eqnarray}
 \rho = r^b , 
\quad 
 u = \frac{(t-c) r - a}{r^b},
\end{eqnarray} 
then the metric \eqref{eq:metric} and the Maxwell field \eqref{eq:field} take the forms,  
\begin{eqnarray}
ds^2 &=&  -\frac{\rho ^2}{H'^2} du^2 + 2 \frac{L}{b H'^2} du d\rho 
+ \frac{H'^3 K - N L^2}{b^2 N H'^2 \rho^2} d\rho ^2
\notag \\ &&
\qquad + \frac{H'K}{N}  d\Omega _{\rm S ^2} ^2 
 + \frac{H' N^3}{ K} (d\psi + \cos \theta d\phi) ^2 ,
\label{eq:metricmirainull}
\\
A_\mu dx^\mu &=&  \pm \frac{\sqrt 3}{2 H' } \left[ \rho du - \frac{L}{b \rho} d\rho \right], 
\label{eq:fieldmirainull}
\end{eqnarray} 
where 
\begin{eqnarray}
 H' = M + \rho ^{1/b}, \quad K=N^2 + a + u \rho + c \rho^{1/b}, 
\quad L=a + (1-b) u \rho. 
\label{eq:metricHK}
\end{eqnarray} 
In the limit $\rho \to 0$ with $u={\rm finite}$, (equivalently, $r \to 0$ with $t r = a$ ), 
the metric \eqref{eq:metricmirainull} and the Maxwell field \eqref{eq:fieldmirainull} behave as 
\begin{eqnarray}
ds^2 &\to & \frac{2 a}{b M^2} du d\rho - \frac{M^4}{4 a^2 b^2 N^2} u^2 d\rho^2
+ M  \frac{ N^2 + a }{N} d\Omega _{\rm S ^2} ^2 
+ \frac{M N^3}{N^2 + a} (d\psi + \cos \theta d\phi) ^2 ,
\label{eq:metricmirainullnearr0}
\\
A_\mu dx^\mu &\to & \pm \frac{\sqrt 3 M^2}{4ab N} u d\rho, 
\label{eq:fieldmirainullnearr0}
\end{eqnarray} 
where a pure gauge term $\rho^{-1}d\rho$ in $A_\mu dx^\mu$ is omitted. 
The metric, which represents AdS$^2\times$(squashed S$^3$), 
and the Maxwell field are regular at $\rho = 0$.  
We also see that the $\rho = 0$ surface is a null surface, 
and the angular part of the metric, which describes a squashed S$^3$, 
does not depend on the time at $\rho = 0$. 
It means that expansion of outgoing null bundle emanating from the squashed S$^3$ on a 
time slice $u={\rm const.}$ is vanishing at $\rho=0$.
The area of the squashed S$^3$ at $\rho=0$ is given by 
\begin{eqnarray}
 A_\text{H} = \sqrt{M ^3 N (N^2 + a)} A_{\rm S ^3} , 
\label{bhmenseki}
\end{eqnarray} 
where $A_{\rm S ^3}$ denotes the area of a unit S$^3$.

\subsection{Extension} \label{extension}

Here, we extend the metric across the $\rho = 0$ surface. 
Similar to the discussion in \cite{Tatsuoka:2011tx}, 
we assume that the function $L$ in \eqref{eq:metricHK} is used globally, 
and the functions $H'$ and $K$ 
in \eqref{eq:metricHK}, which contain $\rho^{1/b}$, are extended as 
\begin{eqnarray}
 H'=M+\Theta(\rho)~ |\rho|^{1/b}, \quad K=N^2 + a + u \rho + c~ \Theta(\rho)~ |\rho|^{1/b},
\quad L=a + (1-b) u \rho, 
\label{eq:extension} 
\end{eqnarray}
where $\Theta(\rho)$ denotes a step function, $\Theta(\rho) = +1 ~(\rho \geq 0),~ 
-1 ~(\rho < 0)$.  
The metric \eqref{eq:metricmirainull} and the Maxwell field \eqref{eq:fieldmirainull} 
with \eqref{eq:extension} 
are continuous at $\rho =0$. 

Introducing new coordinates $r'(< 0)$ and $t'$ in the $\rho<0$ region by
\begin{eqnarray}
 \rho = - (- {r'}) ^b , \quad  u = \frac{(t'-c) (- r') + a}{(-{r'}) ^b} ,
\end{eqnarray} 
we show that 
the metric \eqref{eq:metricmirainull} and the Maxwell field \eqref{eq:fieldmirainull} with \eqref{eq:extension} 
reproduce 
\begin{eqnarray}
 ds^2 &=&  -\left( 1 + \frac{M}{r'} \right)^{-2} dt'^2 
  + \left( 1 + \frac{M}{r'} \right) 
 \left[ \left( \frac{t'}{N} + \frac{N}{r'} \right) ( d{r'}^2 + {r'}^2 d\Omega^2_{\rm S^2} ) 
\right. 
\notag \\ 
&& \qquad \qquad \left. 
+ N^2 \left( \frac{t'}{N} + \frac{N}{r'} \right)^{-1} ( d\psi + \cos \theta d\phi )^2 
\right], 
\label{naibumet}
\\
A_\mu dx^\mu &=&  \pm \frac{ \sqrt{3} }{ 2 } \left( 1 + \frac{M}{r'} \right)^{-1} dt' .
\label{naibumaxwell}
\end{eqnarray}
The metric and the Maxwell field coincide with 
the metric \eqref{eq:metric} and the Maxwell field \eqref{eq:field} 
with $r' < 0$.  
Then it is clear that 
the spacetime with the metric \eqref{eq:metricmirainull} with \eqref{eq:extension} 
gives a C$^0$ extension of the metric \eqref{eq:metric} in the region I to the region II. 
That is, the null boundary $r\to 0_+, t\to \infty$ with $rt=a$ in 
the region I is attached 
the null boundary $r\to 0_-, t\to -\infty$ with $rt=a$ in the region II.

The outer region I becomes asymptotically the Kasner-like universe 
described by \eqref{eq:5dkasner}, 
then it has a null infinity. 
However, any null geodesic starting from a point in 
the inner region II cannot reach the null infinity. Therefore, the $\rho=0$ surface is an event horizon. 
The exact solution \eqref{eq:metric} with \eqref{eq:field} indeed represents 
the charged black hole in the five-dimensional anisotropically expanding Kaluza-Klein universe. 
We see by the metric \eqref{eq:metricmirainullnearr0} 
that the horizon shape is not 
a round S$^3$ but the squashed S$^3$.  
The area of the event horizon is independent of the time.

\subsection{ Penrose diagram} 
\label{globalstructures}
\begin{figure}[h]
\begin{center}
 \includegraphics[width=50mm,clip]{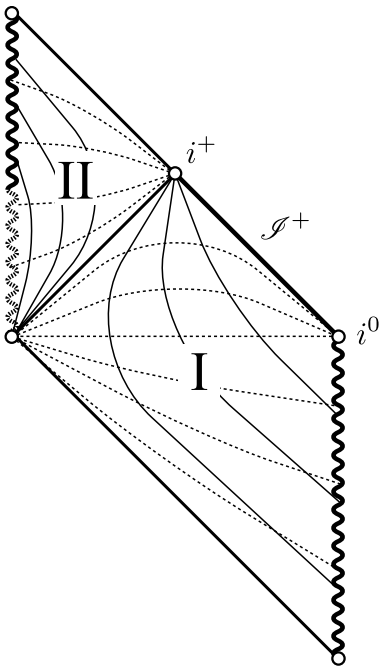}
\end{center}
\caption{Penrose diagram of $t-r$ plane.
The outer region I and the inner region II are joined at 
the event horizon, $r=0$ and $rt=a$. The null infinity exists at $r=+\infty$ and $t=+\infty$. 
The wavy lines are curvature singularities. 
Dashed curves denote $t ={\rm const.}$ surfaces, and 
thin solid curves denote $r ={\rm const.}$ surfaces.}
\label{fig:penrose2}
\end{figure}

According to the extension in the previous subsection, 
the Penrose diagram of the solution \eqref{eq:metric} is shown in 
the Fig.\ref{fig:penrose2}.

In the outer region I,  
the geometry looks like the anisotropic Kasner universe described by \eqref{eq:5dkasner} 
in the far region, $r \gg N,M$, 
where the three-dimensional space expands infinitely, 
while the compact extra dimension shrinks with the time evolution.  
There is a null infinity at $r = +\infty ,~ t = +\infty$. 
There also exists a timelike singularity at $rt = -N^2$. 
In the inner region II, 
there are timelike singularities at $r = -M <0$ and $rt = -N^2$. 
Two regions I and II are attached at the event horizon $\rho=0$ in the new coordinate, that is, at the coordinate 
boundary in the old coordinates 
$r = 0_\pm ,~ t= \pm \infty$ with $tr =a={\rm finite}$.

We have the metric with all of the components are continuous at the event horizon. 
Although the metric is not smooth at the horizon, the Kretschmann invariant is finite  
there, and components of the Ricci tensor are finite in regular coordinate basis, $du$ and $d\rho$. 
We find, however, that some components of Riemann curvature diverge there.  

The singularity on the horizon is relatively mild. 
For example, a component of Riemann curvature diverges at $\rho = 0$ as 
\begin{eqnarray}
 R_{\theta \rho \theta \rho } \propto 
\rho ^{-2+1/b  },  
\end{eqnarray} 
where $1/b$ takes a value in the range $1 < 1/b < 2$.\footnote{
Note that this implies the existence of a parallelly propagated curvature
singularity at the horizon along a null geodesic falling into the black hole
since $\partial/\partial\rho$ is the tangent vector to the approximate null
geodesic that agrees with an exact geodesic at the horizon,
and $\partial/\partial\theta$ is a regular vector at the horizon. }
Because the integration of the curvature component in an infinitesimal segment across the horizon: 
\begin{eqnarray}
 \int_{-\epsilon}^{+\epsilon} d \rho R_{\theta \rho \theta \rho } 
\end{eqnarray} 
is finite, 
the tidal force causes finite difference in deviation of geodesic congruence 
crossing the horizon. 
Then, the singularity is relatively mild so that an observer can traverse the horizon.

We have extended the metric across the surface $r=0$ and $t=\infty$ to clarify   
the spacetime is black hole. 
Although the coordinate boundaries $r=0_+$ and $t=-\infty$ in the region I, 
and $r=0_-$ and $t=+\infty$ in the 
region II would be extendible, explicit extension is not done yet.

\subsection{Static geometry near event horizon}

The spacetime is dynamical because the metric \eqref{eq:metric} does not admit any timelike Killing vector. 
However, \eqref{eq:metricmirainullnearr0} means the size of event horizon, given by \eqref{bhmenseki}, 
is constant during evolution of the 
universe.

To observe the geometry near horizon clearly, 
we consider the limit $r \to 0_+ ,~ t \to \infty$ 
keeping $r t \to {\rm finite}$
of the metric \eqref{eq:metric}. 
In this limit  
the metric has the form of
\begin{eqnarray}
ds^2 \simeq -\frac{r^2}{M^2} dt^2 + \frac{M(rt+N^2)}{N r^2} dr^2 
+ \frac{M(rt+N^2)}{N} d\Omega ^2 _{\rm S ^2} + \frac{M N^3}{rt + N^2} (d\psi +\cos{\theta }d\phi )^2 .
 \label{eq:nearr01}
\end{eqnarray}
Introducing coordinates 
\begin{eqnarray}
 \tilde \rho^2 = \frac{M}{N} (rt+N^2) , 
\quad 
 \tau = \frac{N}{M^2} \log t ,
\end{eqnarray}
we see the metric \eqref{eq:nearr01} becomes 
\begin{eqnarray}
ds^2 &\simeq&  -f(\tilde \rho ) d\tau ^2 - \frac{4 M^2 \tilde \rho ^3}{N (\tilde \rho ^2 - MN)} d\tau d\tilde \rho 
+ \frac{4 \tilde \rho ^4}{(\tilde \rho ^2 - MN)^2} d\tilde \rho ^2
\notag \\
 && \qquad + \tilde \rho ^2 d\Omega ^2 _{\rm S ^2} + \frac{M^2 N^2}{\tilde \rho ^2} (d\psi +\cos{\theta }d\phi )^2 ,
\label{eq:nearr03}
\\
f(\tilde \rho ) &=&  
(\tilde \rho ^2 - \tilde \rho ^2 _+)(\tilde \rho ^2 - \tilde \rho ^2 _-) , 
\quad
\tilde \rho _\pm ^2 = \frac{M (M^3 + 2 N^3) \pm \sqrt{M^5 (M^3 + 4 N^3)}}{2 N^2} . 
\end{eqnarray}
Further, introducing a coordinate 
\begin{eqnarray}
d T = d\tau + \frac{2 M^2 \tilde \rho ^3}{N (\tilde \rho ^2 - MN) f(\tilde \rho )} d\tilde \rho ,
\end{eqnarray}
we have the metric \eqref{eq:nearr03} in the form 
\begin{eqnarray}
ds^2 \simeq -f(\tilde \rho ) d T^2 + \frac{4 \tilde \rho ^4}{f(\tilde \rho )} d\tilde \rho ^2 
+ \tilde \rho ^2 d\Omega ^2 _{\rm S ^2} + \frac{M^2 N^2}{\tilde \rho ^2} (d\psi +\cos{\theta }d\phi )^2 . 
\label{eq:nearr04}
\end{eqnarray}
This metric is a limiting case of static charged squashed Kaluza-Klein black hole solutions derived in 
ref. \cite{Ishihara:2005dp}. Taking the limit that the asymptotic size of the extra dimension 
becomes zero, 
the charged squashed Kaluza-Klein metric reduces to \eqref{eq:nearr04}. 
It is clear that the event horizon is static. 

At the late stage, the three-dimensional distances between observers at 
$r={\rm const.} \neq 0$ and constant angular coordinates increase by the cosmological 
scale factor $\sim \sqrt{t}$, and the size of the extra dimension shrinks 
as $\sim 1/\sqrt{t}$. 
Nevertheless, near black hole region, i.e., $r\to 0,~t\to \infty$ with  $rt={\rm const.}$,   
for an observer at a finite circumference distance $\tilde \rho$ the size of extra dimension 
is the finite time-independent value $MN/\tilde\rho$. 

It is also clear from \eqref{eq:nearr04} that the horizon is non-degenerate 
though the metric is constructed by using the harmonic function $H$ on a Ricci flat 
base space, 
in contrast to stationary extremal charged black holes, which have degenerate horizons.  
This is similar to the cosmological charged black holes with 
a cosmological constant \cite{Ida:2007vi}.

\subsection{Expansion of a null congruence}
\label{nexpansions}

Here we calculate the expansions of the null vector fields   
emanating from the closed surface $r ={\rm const.}$ on a $t ={\rm const.}$ slice.    
The expansions are defined by 
\begin{eqnarray}
\theta ^\pm =h^{\mu \nu} \nabla _\mu k^{(\pm)} _\nu , 
\end{eqnarray}
where 
$k^{(+) \mu} \partial _\mu$ and $k^{(-) \mu} \partial _\mu$ 
denote 
future null vector fields.  
We choose the null vector fields 
$k^{(\pm) \mu} \partial _\mu$
such that 
\begin{eqnarray}
k^{(+) \mu} \frac{\partial}{\partial x ^\mu} &=&  
\frac{1}{\sqrt 2} \frac{\partial}{\partial t} + \frac{1}{H \sqrt{2 H V}} \frac{\partial}{\partial r}  ,
\label{k+}\\
k^{(-) \mu} \frac{\partial}{\partial x ^\mu} &=&   
\frac{H ^2}{\sqrt 2} \frac{\partial}{\partial t} - \sqrt{\frac{H}{2 V}} \frac{\partial}{\partial r} ,
\label{k-}
\end{eqnarray}
where $k^{(\pm)}$ are null vectors in the direction of increasing/decreasing $r$ coordinate, respectively. 
The vectors $k^{(\pm)}$ are regular in the far region, $r \gg M,N$, and they 
satisfy the relations 
$k^{(-) \mu} \nabla _\mu k^{(-) \nu} = 0$ , 
and $g _{\mu \nu} k^{(+) \mu} k^{(-) \nu} = -1$. Then, they are regular everywhere. 
The metric on S$^3$ ($r={\rm const.}$, $t={\rm const.}$), $h_{\mu \nu}$,
is given by 
\begin{eqnarray}
 h_{\mu \nu} = g_{\mu \nu} + k^{(+)} _\mu k^{(-)} _\nu + k^{(+)} _\nu k^{(-)} _\mu. 
\end{eqnarray}
In the case of \eqref{k+} and \eqref{k-}, $h_{\mu \nu}$ becomes 
\begin{eqnarray}
 h_{\mu \nu} dx^\mu dx^\nu = H V r^2 d\Omega^2_{\rm S^2} + N^2 H V^{-1} ( d\psi + \cos \theta d\phi )^2 .
\label{indmetexpand}
\end{eqnarray}

The expansions of null geodesic congruences on the three-dimensional space \eqref{indmetexpand} are obtained as 
\begin{eqnarray}
\theta ^+ &=&  \frac{rH^2\sqrt{HV} + \left( 3N V+tH\right)}{2 \sqrt{2}N r H^{5/2}V^{3/2}} ,
\\
\theta ^- &=&  \frac{rH^2\sqrt{HV} - \left( 3N V+tH\right)}{2 \sqrt{2}N r  H^{1/2}V^{3/2} } .
\end{eqnarray}
In the limit $r \to 0 ,~ t \to \infty$ with $tr=a$, 
$\theta ^+ = 0$ and $\theta ^- ={\rm const.}< 0$.  
Then we see that the event horizon, $r = 0 ,~ t = \infty$ with $tr = a$ surface, is an apparent horizon.

\begin{figure}[!h]
\includegraphics[width=65mm]{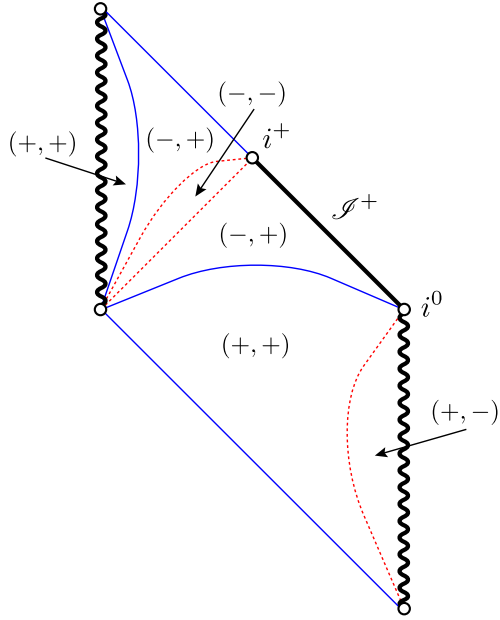}
\caption{
The sign of $\theta ^\pm$ in the Penrose diagram.
Pairs of $(\pm, \pm)$ denote the sign of $(\theta ^- , \theta ^+)$.   
Dotted (red) curves and solid (blue) curves denote 
$\theta ^+ = 0$ and $\theta ^- = 0$ 
surfaces, respectively. 
}
\label{fig:penrose4}
\end{figure}

We show the sign of $\theta ^\pm $ in the Penrose diagrams in the Fig.\ref{fig:penrose4}.  
We can see that 
outside the black hole, $r > 0$, 
there is a region of $(\theta^+ ,\theta ^- )=(+,+)$ like an expanding universe, 
while inside the black hole, $r < 0$,  
there is a trapped region $(\theta^+ ,\theta ^- )=(-,-)$ .  

\vspace{1cm}

\section{Summary} \label{summary}

We examine the exact solution which represents a charged black hole 
resides in a five-dimensional Kasner-like universe 
in the Einstein-Maxwell theory. 
Outside of the black hole horizon, $r>0$, the metric approaches to the five-dimensional 
Kasner-like universe, where three-dimensional space expands while a compact 
extra dimension shrinks, in the far region. 
The universe has a future null infinity in the late time and a timelike singularity 
in the early stage. 
Inside of the black hole horizon, $r<0$, there are also timelike singularities. 
We give a C$^0$ extension of the spacetime across the event horizon $r=0$. 
Thus the solution represents the charged black hole sitting in the dynamical 
Kaluza-Klein universe.  
The shape of the horizon is a squashed S$^3$, 
and its area does not depend on the time though the spacetime is dynamical.  


The metric is not smooth at the horizon. Even though the Kretschmann invariant is finite, 
some components of Riemann curvature in a regular basis diverge at the horizon. 
The curvature singularity is relatively mild because an integration of the curvature 
in an infinitesimal segment across the horizon is finite. 
Therefore, a free-falling observer can traverse the horizon. 
In contrast to the fact that  
the five-dimensional stationary squashed Kaluza-Klein black hole solutions 
\cite{Dobiasch:1981vh, Gibbons:1985ac, Gauntlett:2002nw, 
Gaiotto:2005gf, Ishihara:2005dp, Wang:2006nw, 
Yazadjiev:2006iv, Nakagawa:2008rm, Tomizawa:2008hw, Tomizawa:2008rh, Stelea:2008tt,  
Tomizawa:2008qr, Gal'tsov:2008sh, Matsuno:2008fn, Bena:2009ev, Tomizawa:2010xq, 
Mizoguchi:2011zj, Chen:2010ih, Stelea:2011fj, Nedkova:2011hx,  
Tatsuoka:2011tx, Nedkova:2011aa, Mizoguchi:2012vg, 
Matsuno:2012hf, Matsuno:2012ge, Tomizawa:2012nk, Stelea:2012ph, Nedkova:2012yn, Wu:2013nea, 
Brihaye:2013vsa} 
have smooth horizons,    
the five-dimensional dynamical Kaluza-Klein black hole has weakly singular event horizon. 
This is similar to the extremal charged Kaluza-Klein black hole 
solutions \cite{Tatsuoka:2011tx} in the case of higher than five dimensions. 


Numbers of extremal charged stationary black hole solutions are constructed 
by using harmonic functions on Ricci flat base spaces. 
In these cases, black hole horizons are degenerate. 
Similarly, the solution in the present paper is constructed by a harmonic function on a Ricci flat 
base space. However, the base space in the present case has the time-variable as a parameter. 
This is the reason why the solution is dynamical. Resultant black hole has non-degenerate horizon 
in this case. It is worth noting that the set of metric and Maxwell field is also a solution of 
the Einstein-Maxwell-Chern-Simons system. 

In the late time, though the size of extra dimension in the far region shrinks to a smaller size 
than that of the black hole, the event horizon does not change in its size. 
Indeed, the total spacetime is dynamical, but the geometry of near the event horizon is static.  
The expansion of the outgoing null geodesic congruence emanating from the event horizon is vanishing, i.e., 
the event horizon is an apparent horizon even though the spacetime is dynamical. 
We can understand this result as follows.
In static vacuum Kaluza-Klein black hole solutions~\cite{Dobiasch:1981vh, Gibbons:1985ac}, 
the horizons are flattened as the size parameters of the extra dimension become small. 
In contrast, in the case of charged Kaluza-Klein black holes~\cite{Ishihara:2005dp}, 
the horizon can be fat and round against to the small extra dimension. 
Thus, we would expect that the existence of electric charge stabilizes the size of black hole 
against shrinking extra dimension of the Kaluza-Klein universe in the present solution.

The solution \eqref{eq:metric} can be easily generalized to multi-black hole solution. 
In this solution, 
the metric \eqref{eq:metric}, 
the harmonic functions \eqref{eq:h} and \eqref{eq:v} are replaced by 
\begin{eqnarray}
 ds^2 &=&  -H^{-2} dt^2 + H \left[ V ( dx^2 + dy^2 + dz^2 ) + V^{-1} ( d\zeta + \bm \omega )^2 \right] , 
 \label{eq:metric2}
\\ 
 H &=&  1 + \sum _i \frac{ M_i }{ |\bm x - \bm x _i| } ,
 \label{eq:h2}
\\ 
 V &=&  \frac{ t }{ t_0} + \sum _i \frac{ N_i }{ |\bm x - \bm x _i| } ,
 \label{eq:v2}
\end{eqnarray}
where the 1-form $\bm \omega $ is determined by $\nabla \times \bm \omega = \nabla V$, 
and $t_0,~ M_i ,~N_i$ are positive constants, and 
$\bm x = (x ,~ y ,~ z) ,~ \bm x_i = (x_i ,~ y_i ,~ z_i)$ denotes position vectors on the three-dimensional 
Euclid space. 
The metric \eqref{eq:metric2} with the harmonic functions \eqref{eq:h2} and \eqref{eq:v2}  
would describe the 
multi-black holes. 
We leave the analysis for the future.

\section*{Acknowledgments}
We would like to thank M. Nozawa for fruitful comments and suggestions. 
We also thank T. Houri, K.-i. Nakao, Y. Yasui, and C.-M. Yoo for useful discussions. 
This work is supported by the Grant-in-Aid for Scientific Research No.19540305 
and 24540282. 
M.K. is supported by a grant for research abroad from JSPS.



\end{document}